\documentclass{PoS}

\input pix.sty

\usepackage{graphicx}
\usepackage{epstopdf}

\newcommand{\sumint}[1]{\mbox{$\sum$}\!\!\!\!\!\!\!\int_{#1}}

\newcommand{\dd}{\mathrm{d}}
\newcommand{\tinymsbar}{{\overline{\mbox{\tiny\rm{MS}}}}}
\newcommand{\Lambdamsbar}{{\Lambda_\tinymsbar}}

\newcommand{\Nc}{N_{\rm c}}

\newcommand{\gB}{g_\rmii{B}}

\newcommand{\bmu}{\bar\Lambda} 

\def\lsi{\raise0.3ex\hbox{$<$\kern-0.75em\raise-1.1ex\hbox{$\sim$}}}
\def\gsi{\raise0.3ex\hbox{$>$\kern-0.75em\raise-1.1ex\hbox{$\sim$}}}

\newcommand{\rmii}[1]{{\mbox{\tiny\rm{#1}}}}

\newcommand{\im}{\mathop{\mbox{Im}}}

\newcommand{\Tint}[1]{{\hbox{$\sum$}\!\!\!\!\!\!\!\int\,}_{\!\!\!\!\raise-0.9ex\hbox{$\scriptstyle{#1}$}}}
\newcommand{\Tinti}[1]{{{\Sigma}\!\!\!\!\raise0.3ex\hbox{$\int$}_\rmii{${#1}$}}}
\renewcommand{\Tint}[1]{\sumint{#1}}

\newcommand{\bi}{\begin{itemize}}
\newcommand{\ei}{\end{itemize}}

\newcommand{\hide}[1]{ }

\title{Bulk and shear spectral functions in weakly and strongly coupled Yang-Mills theory}

\ShortTitle{Spectral functions in Yang-Mills theory}

\author{\speaker{Aleksi Vuorinen}\\
        Bielefeld University\\
        E-mail: \email{vuorinen@physik.uni-bielefeld.de}}

\author{Martin Kr\v{s}\v{s}\'ak\\
        Bielefeld University\\
        E-mail: \email{krssak@physik.uni-bielefeld.de}}

 \author{Yan Zhu\\
         Bielefeld University\\
         E-mail: \email{yzhu@physik.uni-bielefeld.de}}        
        
\abstract{In this talk, we discuss a number of recent calculations aimed at determining the spectral functions corresponding to various components of the energy momentum tensor in high-temperature SU($N$) Yang-Mills theory. The computations reviewed include applications of both weak coupling and gauge/gravity techniques, and thus enable one to access different limits of the quantities. The motivation for the work is twofold: On one hand, the results are hoped to aid the eventual nonperturbative extraction of the bulk and shear viscosities from lattice data, while on the other hand they also enable an immediate comparison of the lattice, perturbative and holographic predictions for certain Euclidean correlators.}

\FullConference{Xth Quark Confinement and the Hadron Spectrum,\\
		October 8-12, 2012\\
		TUM Campus Garching, Munich, Germany}

\begin{document}

\section{Motivation and setup}

The past decade has witnessed a remarkable progress in the hydrodynamic description of the heavy ion experiments conducted at RHIC and the LHC \cite{Tannenbaum,Muller}. In addition to providing a more coherent picture of the evolution of the expanding fireball, this development has also highlighted the importance of having a quantitatively accurate handle on the transport properties of the quark gluon plasma (QGP). This necessitates in particular the development of new tools to predict the values of the transport coefficients that appear in the hydrodynamic equations governing the dynamics of the system, of which the shear viscosity is perhaps the most prominent example \cite{Romatschke:2009im,Shen:2011zc}. An analysis of the elliptic flow data from RHIC has namely pointed towards a small but nonvanishing value for the shear viscosity to entropy ratio $\eta/s$, which is in clear contrast with perturbative expectations \cite{Aarts:2002cc,Arnold:2003zc}, but in surprisingly good agreement with the value obtained for strongly coupled large-$N_c$ field theories via the gauge/gravity duality \cite{Kovtun:2004de} (see also \cite{Kovtun:2011np,Rebhan:2011vd} for some recent developments). This has in particular led to the famous conjecture of the QGP produced at RHIC being a nearly `ideal' fluid; at LHC, things may, however, be markedly different.

Unfortunately, the first principles determination of transport coefficients in an interacting quantum field theory is a complicated problem. In the weakly coupled limit of asymptotically high temperatures, only low order perturbative results are available, and their convergence appears to be far from optimal (see e.g.~\cite{Arnold:2003zc} and references therein). Closer to the critical temperature $T_c$, one clearly needs nonperturbative machinery, but due to the restriction of lattice QCD to the Euclidean formulation of the theory, the transport coefficients --- obtainable from the IR limit of the corresponding Minkowskian spectral functions --- are not directly available \cite{Meyer:2007ic,Meyer:2007dy,Meyer:2011gj,Brandt:2012jc}. Finally, in the infinitely strongly coupled limit of a class of (mostly conformal) theories \cite{Huebner:2008as,Iqbal:2009xz,Springer:2010mf,Springer:2010mw,Kajantie:2010nx,Kajantie:2011nx}, for which five-dimensional gravity duals exist, both spectral functions and transport coefficients are available via relatively straightforward calculations. In this context, a nontrivial question however is, how to relate the dual field theories to real life QCD; to this end, a particularly promising direction has been the development of Improved Holographic QCD (IHQCD) \cite{Gursoy:2007cb} --- a five-dimensional gravity-dilaton system designed to systematically mimic the properties of quarkless QCD both in the IR and UV (see also \cite{Jarvinen:2011qe,Alho:2012mh} for recent work towards including fermionic effects in the model).

In this talk, our goal is to review and discuss the results of a series of recent works, in which the task of determining the bulk and shear channel spectral functions in SU($N$) Yang-Mills theory has been undertaken  \cite{Kajantie:2011nx,Laine:2011xm,Zhu:2012be,Krssak}. The motivation for this comes from two separate directions. Due to asymptotic freedom, perturbation theory is expected to provide an accurate description of the UV behavior of various correlators. This makes it a vital ingredient in any attempt to perform an analytic continuation of Euclidean lattice data to Minkowskian signature \cite{Burnier:2011jq}, necessary to obtain nonperturbative first principles predictions for the corresponding transport coefficients.\footnote{For two successful examples of this, see \cite{Burnier:2012ts,Burnier:2012ze}.} At the same time, the spectral functions (as well as the Euclidean correlation functions that can be determined from them) are also interesting quantities as such. A comparison between the lattice, perturbative and holographic predictions for the correlators namely allows one to assess the extent, to which QGP can be characterized as being `weakly' or `strongly' coupled at various temperatures, constituting one of the most prominent open questions in the field. 

\begin{figure}[t]
\centering
\includegraphics[width=6.1cm]{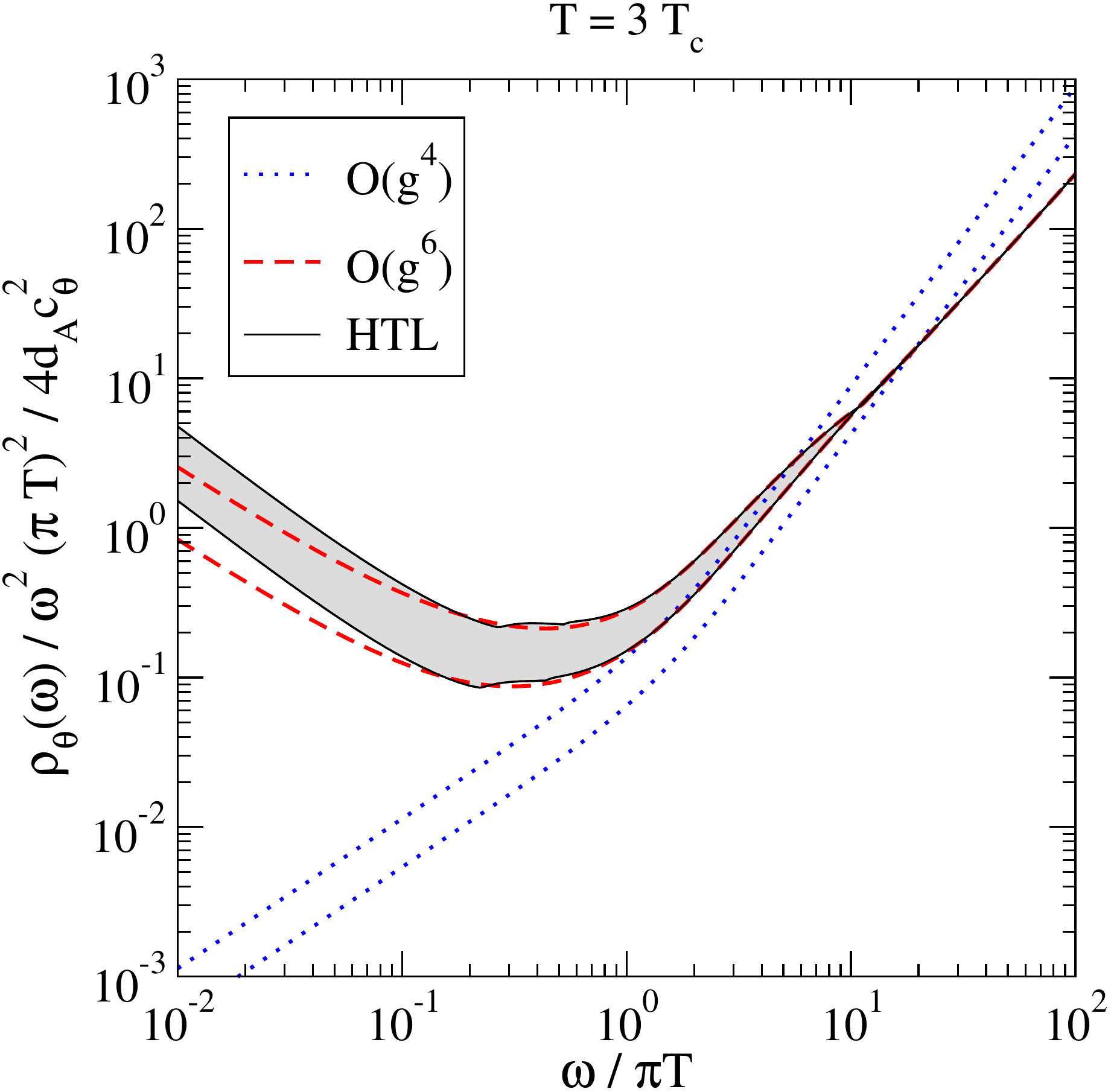}$\;\;\;\;\;\;\;\;\;$\includegraphics[width=8.4cm]{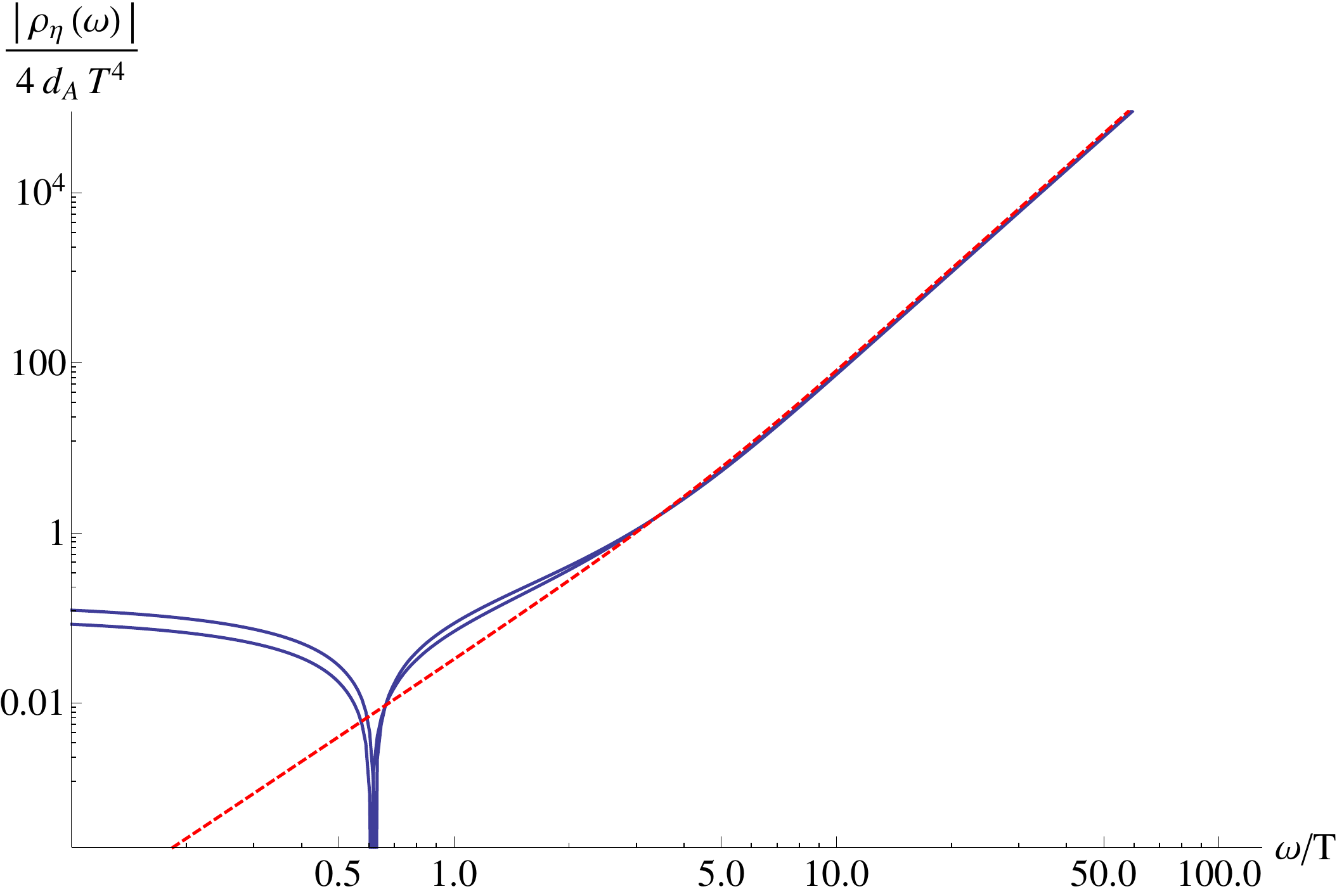}
\caption{The bulk \cite{Laine:2011xm} and shear \cite{Zhu:2012be} channel spectral functions of SU(3) Yang-Mills theory, shown on a logarithmic scale for $T=3T_c$ or 3.75$\Lambdamsbar$. In the latter, the red dashed curve corresponds to the LO result, while the blue curves stand for the NLO ones. The spike in the NLO shear spectral function corresponds to the sign of the quantity turning from positive to negative (with increasing $\omega$).}
\label{fig1}
\end{figure}

We close the introductory part by briefly summarizing the setting of the works we cover, explained in more detail e.g.~in \cite{Laine:2011xm,Zhu:2012be}. Throughout our presentation, we work within pure SU($N$) Yang-Mills theory at a nonzero temperature $T$, defined by the Euclidean action
\ba
 S_\mathrm{E} &=& \int_{0}^{\beta} \! \dd \tau \int \! {\rm d}^{3-2\epsilon}\vec{x}
 \, \frac{1}{4} F^a_{\mu\nu} F^a_{\mu\nu} \; \equiv \; \int_x \, \frac{1}{4} F^a_{\mu\nu} F^a_{\mu\nu} \,, \label{action}
\ea
where we have written $\beta\equiv 1/T$ and
\ba
F^a_{\mu\nu} &\equiv&  \partial_\mu A^a_\nu - \partial_\nu A^a_\mu + \gB f^{abc} A^b_\mu A^c_\nu\, .
\ea
The energy-momentum tensor of the theory takes the form
\ba\la{eq:T}
 T_{\mu\nu} &=& \frac{1}{4} \delta_{\mu\nu} F^a_{\alpha\beta} F^a_{\alpha\beta} -F^a_{\mu\alpha} F^a_{\nu\alpha} \, ,
\ea
which helps us define the bulk and shear operators
\ba
T_{\mu\mu}&\equiv& \theta\, = \, \frac{\beta(g)}{2g} F^a_{\mu\nu}F^a_{\mu\nu}, \\
T_{12}&\equiv& \eta/(4i)\, ,
\ea
where the normalization of the latter has been chosen to be in accordance with \cite{Zhu:2012be}. For both of these operators, we define the spectral function through
\be
 \rho_X (\omega) 
 = \im \bigl[ \tilde G_X(\omega,0) 
 \bigr]
 \;, \la{rho_general}
\ee
where $\tilde G_X(p_0,\mathbf{p})$ is the corresponding momentum space retarded Green's function. It is important to note that to simplify our task, we have here set the external three-momentum $\mathbf{p}$ to zero here.

\section{Perturbation theory \label{pert}}

While leading order (i.e.~noninteracting) spectral functions are rather straightforward quantities to determine, it is only very recently that the first next-to-leading order corrections to them were evaluated in a thermal quantum field theory \cite{Laine:2011xm}. This work concentrated on the bulk channel of Yang-Mills theory, where --- building on the earlier analysis of the UV limit of the quantity in \cite{Laine:2010tc} --- it developed a machinery, with which NLO spectral functions can be systematically reduced to sums of numerical one- and two-dimensional phase space integrals. In \cite{Zhu:2012be,Schroder:2011ht}, this method was subsequently generalized to the technically more complicated case of the shear channel, in which the integrands are no longer rotationally invariant in momentum space.

\begin{figure}[t]
\centering
\includegraphics[width=6cm]{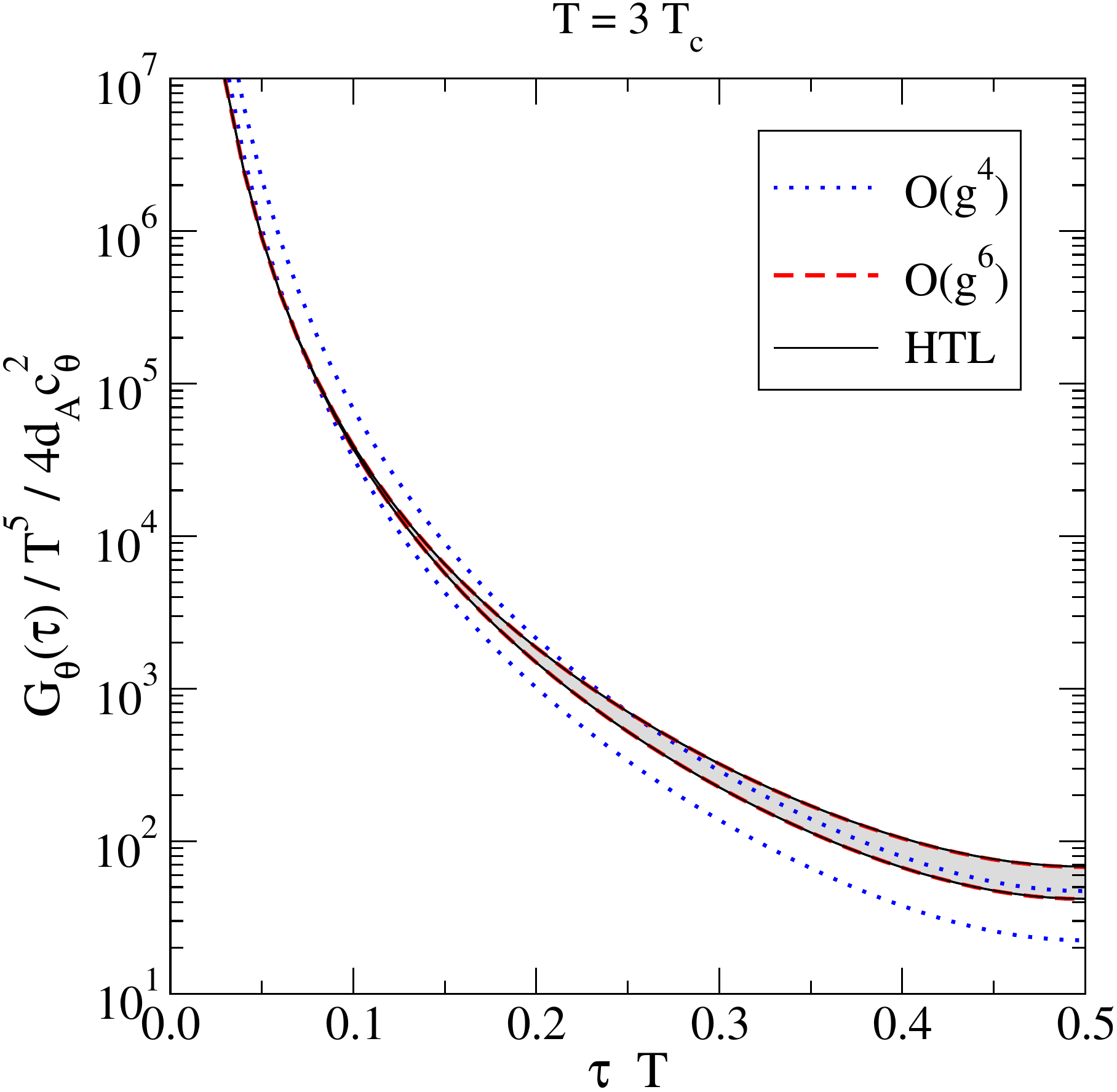}$\;\;\;\;\;\;\;\;\;$\includegraphics[width=8.5cm]{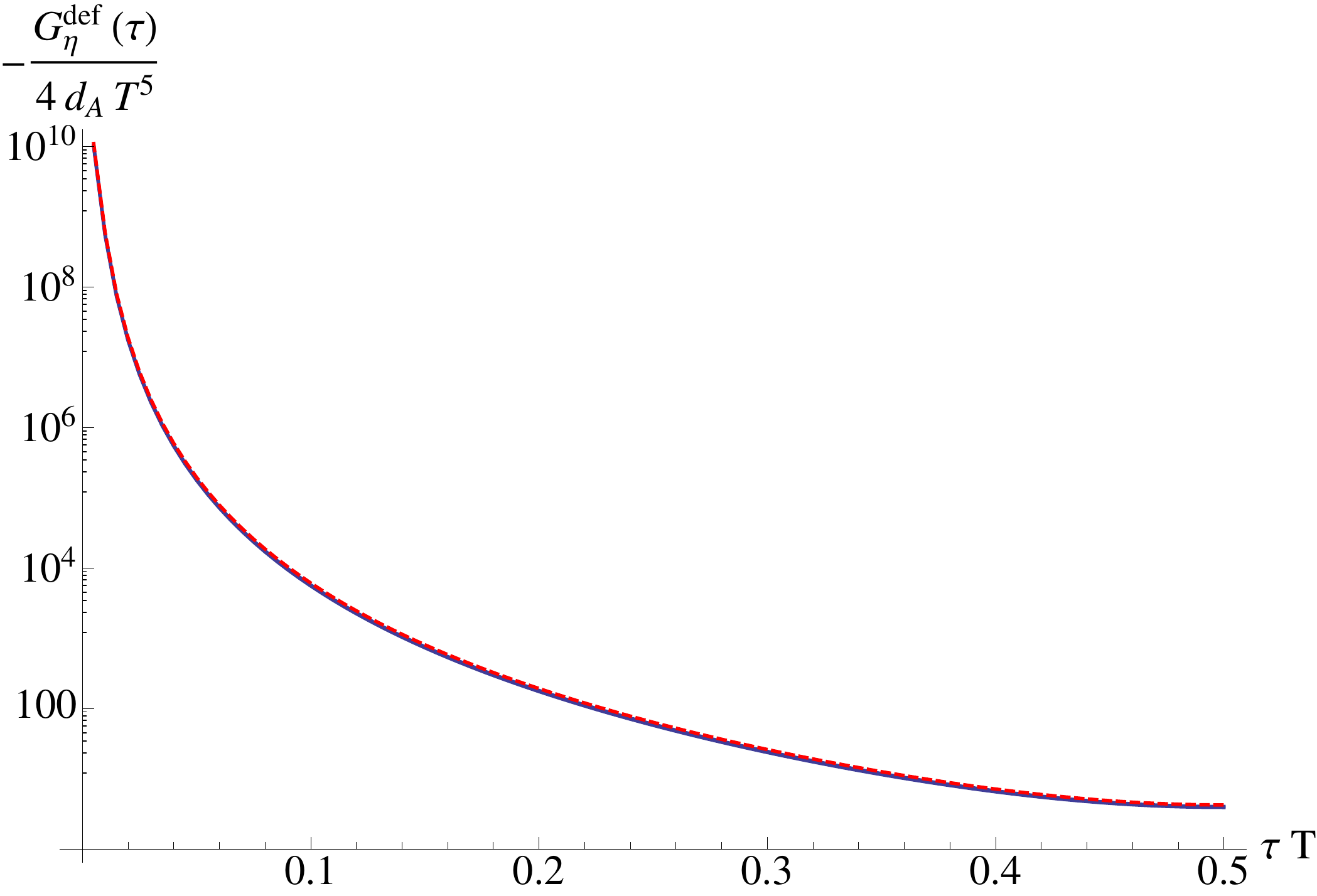}
\caption{The imaginary time correlators in the bulk and shear channels of SU(3) Yang-Mills theory, evaluated at $T=3T_c$ \cite{Laine:2011xm,Zhu:2012be}. In the shear plot, the red dashed line again corresponds to the LO and the blue curves to the NLO result.}
\label{fig2}
\end{figure}

The main results of \cite{Laine:2011xm,Zhu:2012be} have the forms
\ba
 \frac{\rho^{ }_\theta(\omega)}{4 d_A c_\theta^2 } & = &  
 \frac{\omega^4}{16\pi}
 \bigl( 1 + 2 n_{\frac{\omega}{2}} \bigr)
 \biggl\{
   g^4 + \frac{g^6\Nc}{(4\pi)^2}
    \biggl[
       \frac{22}{3} \ln\frac{\bmu^2}{\omega^2} + \frac{73}{3} 
     + 8\, \phi_T^\theta(\omega/T) 
    \biggr] 
 \biggr\}
 + {\mathcal O}(g^8)
 \;, \la{rhofinal_theta} \\
\frac{\rho_\eta(\omega)}{4d_A}&=&\frac{\omega^4}{4\pi}\bigl( 1 + 2 n_{\frac{\omega}{2}} \bigr)\bigg\{-\frac{1}{10}+\frac{g^2N_c}{(4\pi)^2}\bigg[\frac{2}{9}+\phi_T^\eta(\omega/T)\bigg]\bigg\} + {\mathcal O}(g^4)\; , \label{result1}
\ea
where $d_A\equiv N_c^2-1$, $c_\theta= 11N_c/(6(4\pi)^2) + {\mathcal O}(g^2)$ is a normalization constant, and the functions $\phi_T(\omega/T)$ must be determined numerically. The behavior of these quantities is displayed for the case of $N_c=3$ in fig.~\ref{fig1}, with the renormalization scale varied by a factor of 2 around an optimal value (see \cite{Laine:2011xm,Zhu:2012be} for details). For the case of the bulk channel, we have furthermore performed a Hard Thermal Loop type resummation for frequencies $\omega= {\mathcal O} (gT)$ and smaller, producing a visible modification of the IR behavior of the quantity.

As can be seen from the results, convergence in both channels is good as long as $\omega\gtrsim T$, being particularly impressive in the shear case. This can be attributed to the small relative size of the NLO $T=0$ correction\footnote{See also \cite{Zoller:2012qv} for the next few orders in the zero-temperature result.} as well as to the fact that the leading $T$-dependent terms contributing to the function $\phi_T^\eta(\omega/T)$ vanish. In fact, it can be shown that in the $\omega\to\infty$ limit this function behaves as ${\mathcal O}(T^6/\omega^6)$, consistently with the arguments of \cite{CaronHuot:2009ns}. For smaller values of $\omega$, the convergence of perturbation theory clearly deteriorates, as is evident from the large relative size of the NLO corrections. In the case of the shear channel, the spectral function even changes its sign at $\omega\approx 0.6 T$, moving from negative to positive values; this clearly signifies the complete breakdown of our weak coupling expansion.

Two natural applications of our results can be found in sum rules \cite{Romatschke:2009ng,Meyer:2010gu} and Euclidean imaginary time correlators. Concentrating on the latter, we insert the bulk and shear spectral functions into the relation
\begin{equation}
 G(\tau) =
 \int_0^\infty
 \frac{{\rm d}\omega}{\pi} \rho(\omega)
 \frac{\cosh\Big[\! \left(\frac{\beta}{2} - \tau\right)\omega\Big]}
 {\sinh\frac{\beta \omega}{2}}\, ,\quad \quad 0<\tau <\beta \, ,  \la{int_rel}
\end{equation}
obtaining the quantities shown in fig.~\ref{fig2}. Again, we see a remarkable degree of convergence in particular in the shear case, where the LO and NLO curves practically overlap. In the bulk channel, we may further compare the result to the lattice data of \cite{Meyer:2010ii}, demonstrating the cancelation of the UV (small $\tau$) divergence in the difference of the two quantities. This, in fact, is the very reason why perturbative calculations are hoped to be of use in the analytic continuation of Euclidean lattice data to Minkowskian signature, cf.~e.g.~\cite{Burnier:2011jq}.

\begin{figure}[t]
\centering
\includegraphics[width=6cm]{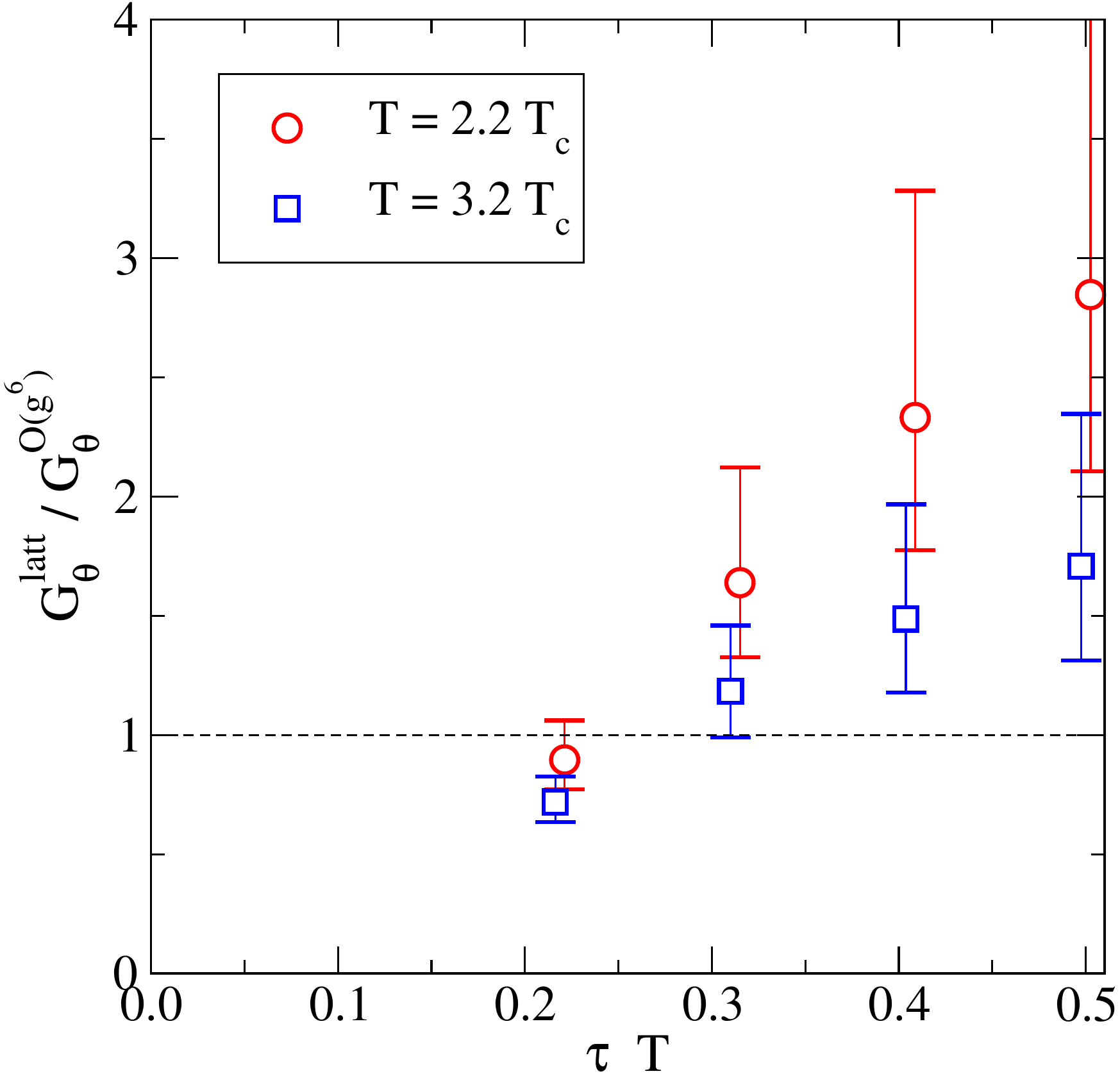}$\;\;\;\;\;\;\;\;\;$\includegraphics[width=6.3cm]{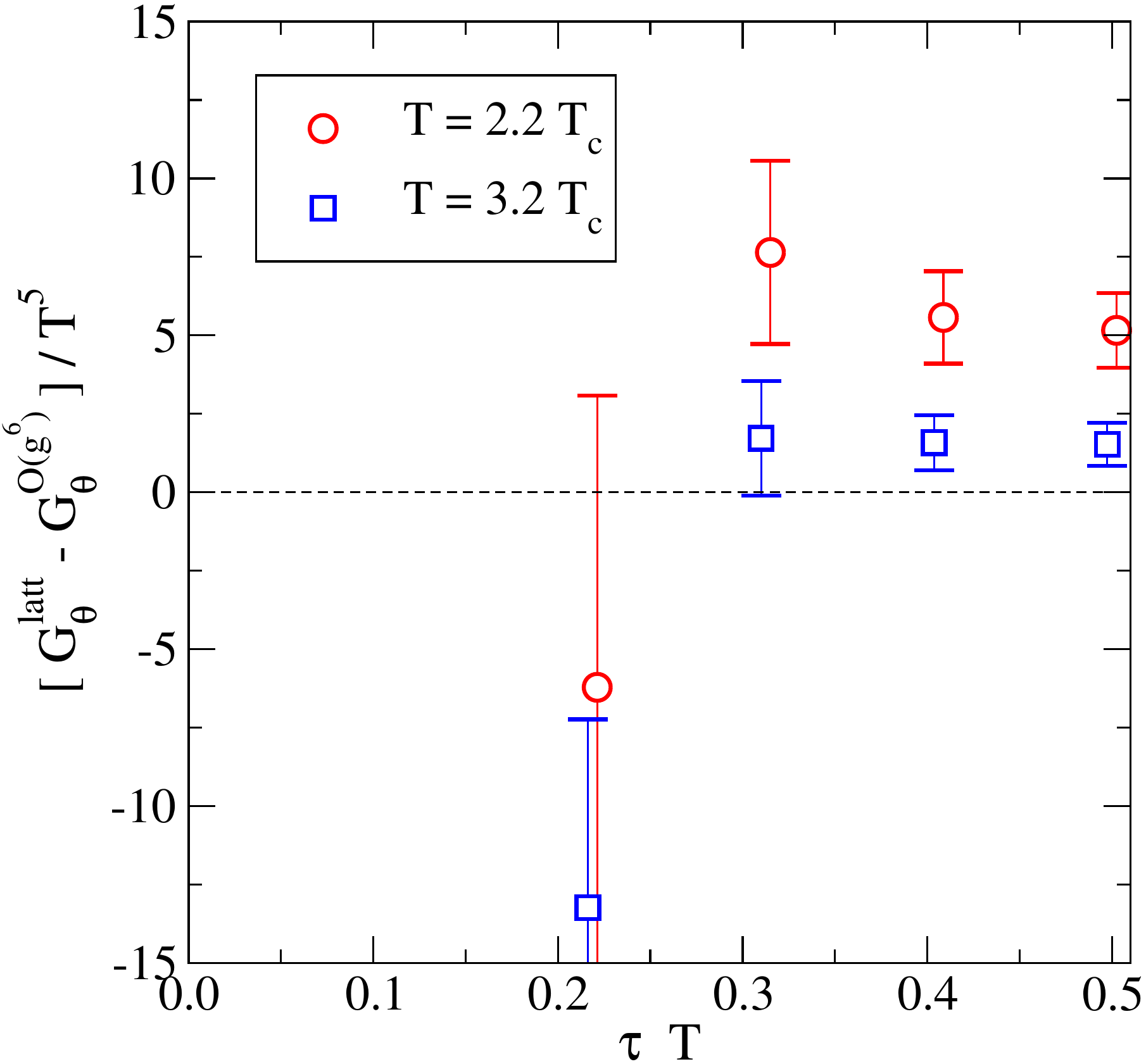}
\caption{A comparison of the ${\mathcal O}(g^6)$ perturbative and lattice results for the imaginary time correlator in the bulk channel of SU(3) Yang-Mills theory.}
\label{fig3}
\end{figure}

\section{AdS/CFT duality and IHQCD \label{ads}}

Moving on to the strong coupling side, we note that spectral functions corresponding to various components of the energy momentum tensor of ${\mathcal N}=4$ Super Yang-Mills theory and its variants have been considered using holographic methods in several works, including e.g.~\cite{Huebner:2008as,Iqbal:2009xz,Springer:2010mf,Springer:2010mw,Kajantie:2010nx,Kajantie:2011nx}. Here, we restrict our attention to one specific five-dimensional model conjectured to be dual to nonsupersymmetric large-$N_c$ Yang-Mills theory, the gravity/dilaton system IHQCD \cite{Gursoy:2007cb}. Within this model, the shear channel spectral function was considered recently in \cite{Kajantie:2011nx}, while the bulk channel will be covered in a forthcoming paper \cite{Krssak}.

\begin{figure}[t]
\centering
\includegraphics[width=7cm]{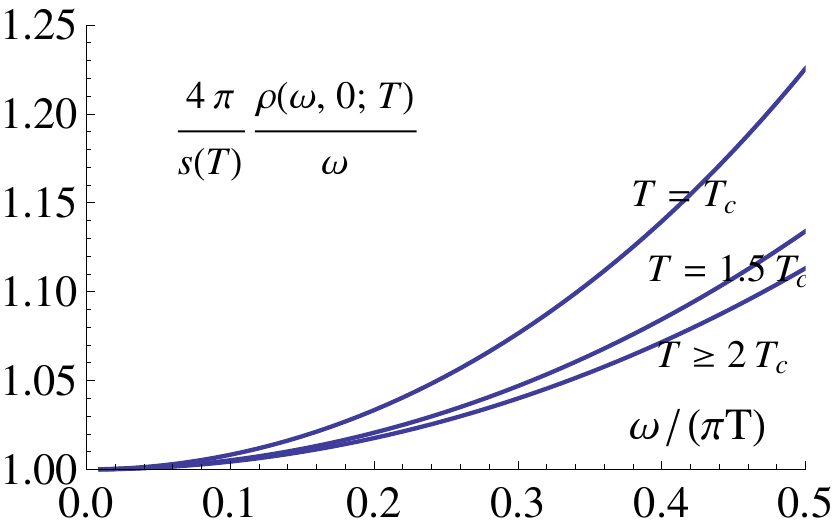}$\;\;\;\;\;\;\;\;\;$\includegraphics[width=7cm]{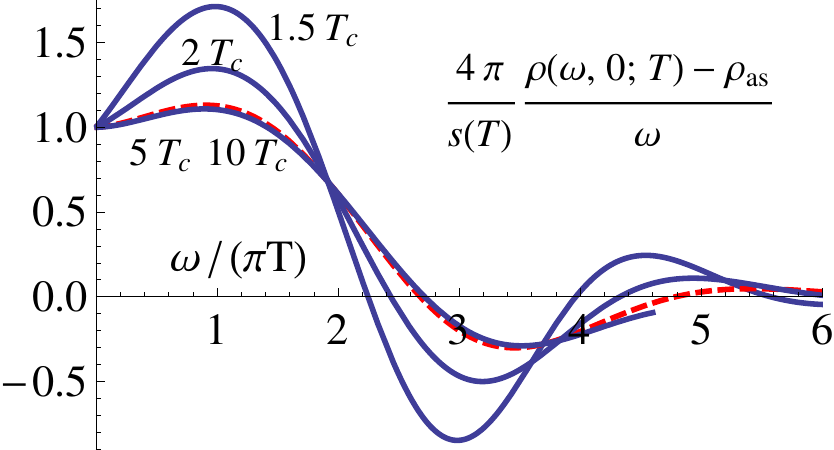}
\caption{The shear channel spectral function as a function of $\omega$, evaluated for a variety of temperatures in the holographic IHQCD model. On the right, the asymptotic (T=0) limit has been subtracted from the quantity, revealing a more complicated nonmonotonous behavior. See \cite{Kajantie:2011nx} for more details.}
\label{fig4}
\end{figure}

Beginning from the shear case of \cite{Kajantie:2011nx}, we display in fig.~\ref{fig4} the behavior of the spectral function for various values of the temperature, given in units of the IHQCD deconfinement transition temperature $T_c$. As is customary in holographic calculations, the evaluation of the retarded correlator reduces to solving a fluctuation equation for the field dual to the boundary operator in question. For the energy momentum tensor, the dual fields are components of the five-dimensional metric tensor, in this case its 12 component. As can be seen from the left part of fig.~\ref{fig4}, the spectral function automatically reproduces the famous holographic prediction $\eta/s = 1/(4\pi)$, while on the right we close in on the temperature dependent part of $\rho$. It is interesting to contrast these predictions with the corresponding results obtained for the conformal ${\mathcal N}=4$ SYM theory in \cite{Kajantie:2010nx} as well for weakly coupled SU($N$) Yang-Mills theory in \cite{Zhu:2012be}; a detailed account of these issues will appear soon \cite{Krssak}.

\begin{figure}[t]
\centering
\includegraphics[width=8.3cm]{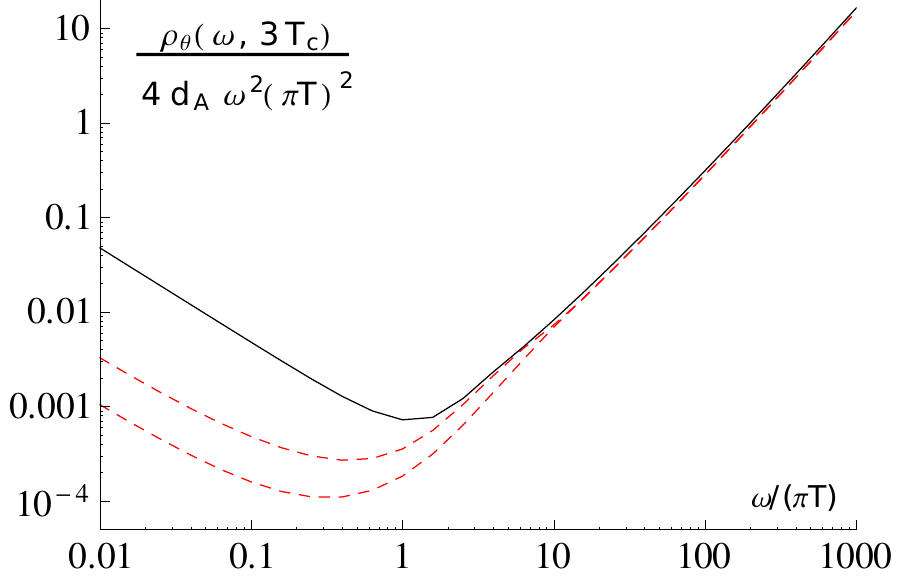}$\;\;\;\;\;\;\;\;\;$\includegraphics[width=5.9cm]{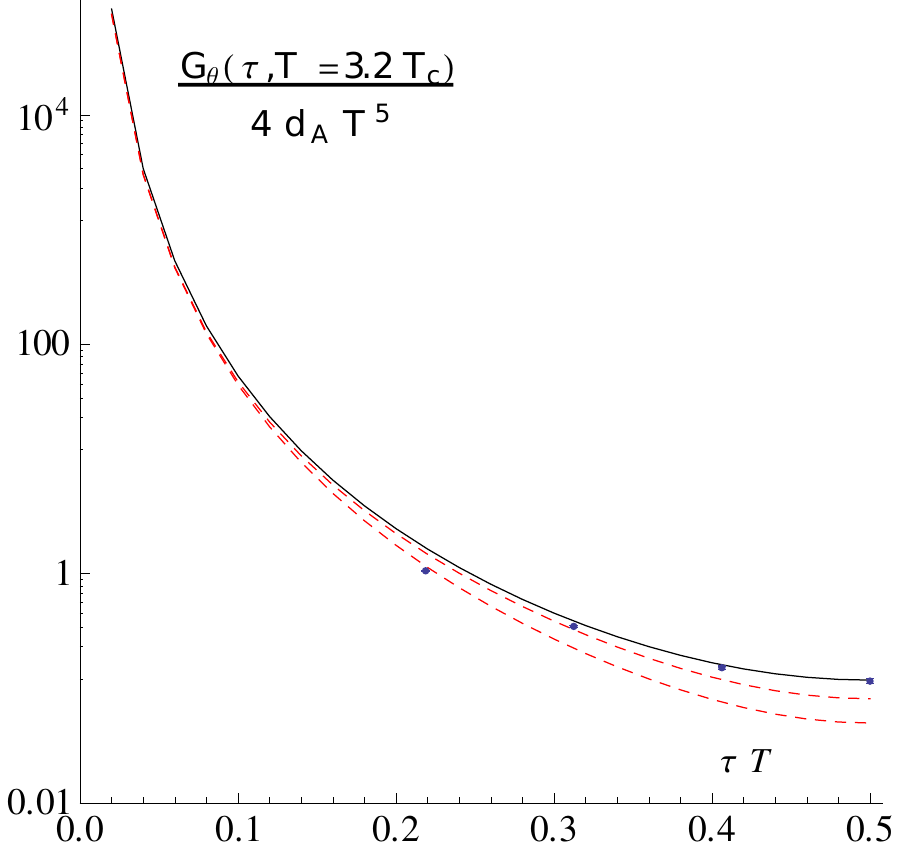}
\caption{Left: A comparison of the IHQCD (black curve) and NLO perturbative (red dashed curves) predictions for the bulk channel spectral function, with the former corresponding to the large-$N_c$ limit and the latter to $N_c=3$. Right: A similar comparison of the imaginary time correlator, with the blue dots corresponding to the $N_c=3$ lattice data of \cite{Meyer:2010ii}.}
\label{fig5}
\end{figure}

Moving finally on to the bulk spectral function, the unpublished results of \cite{Krssak} show impressive agreement with not only the UV limit of the perturbative spectral function in SU(3) Yang-Mills theory, but also the Euclidean imaginary time correlator evaluated on the lattice \cite{Meyer:2010ii}. This is demonstrated in fig.~\ref{fig5}, where we first plot the spectral function on a logarithmic scale and next to it the imaginary time correlator. Two details to note are the automatic agreement of the UV limits in the first plot (i.e.~without any rescalings of the quantity), as well as the remarkable agreement of the three largest datapoints in the second. It should be highly interesting to extend this comparison to a wider range of temperatures to test the robustness of the agreement.

\section{Conclusions and future directions}

In this talk, we have presented and discussed a set of recent results concerning the behavior of the bulk and shear channel spectral functions in thermal SU($N$) Yang-Mills theory. These quantities are crucial ingredients in the description of a (near-)equilibrium quark gluon plasma, as they allow the determination of the corresponding transport coefficients, the bulk and shear viscosities. Unfortunately, the nonperturbative evaluation of these functions via lattice QCD is a notoriously challenging problem, and to this end, any input one can gather via weak coupling or gauge/gravity methods is extremely valuable. In particular, perturbation theory may turn out to be of direct use in the analytic continuation of Euclidean lattice data to Minkowskian signature, as recently demonstrated in \cite{Burnier:2011jq,Burnier:2012ts,Burnier:2012ze}. Our hope is indeed that our perturbative spectral functions, derived in \cite{Laine:2011xm,Zhu:2012be} and discussed in the above sec.~\ref{pert}, will turn out to be 
useful in this process.

While some important developments have recently been reached within both the lattice and perturbative fronts, we are currently still some way from obtaining accurate first principles results for the QGP transport coefficients in a temperature range relevant for heavy ion physics. In the meantime, the results reported in this talk can, however, also be used to evaluate various Euclidean correlators, facilitating a direct comparison between lattice, perturbation theory and gauge/gravity methods. As demonstrated in sec.~\ref{ads} and discussed in more detail in a forthcoming paper \cite{Krssak}, this offers a new and interesting tool for addressing such questions as the weakly/strongly coupled nature of the QGP near the deconfinement transition.

\begin{thebibliography}{99}

\bibitem{Tannenbaum}
  M.~J.~Tannenbaum,
  arXiv:1201.5900 [nucl-ex].

\bibitem{Muller}
  B.~Muller, J.~Schukraft and B.~Wyslouch,
  arXiv:1202.3233 [hep-ex].


\bibitem{Romatschke:2009im}
  P.~Romatschke,
  Int.\ J.\ Mod.\ Phys.\  {\bf E19 } (2010)  1--53,
  [arXiv:0902.3663].

\bibitem{Shen:2011zc}
  C.~Shen, S.~Bass, T.~Hirano, P.~Huovinen, Z.~Qiu, H.~Song, U.~Heinz,
  [arXiv:1106.6350].
  
\bibitem{Aarts:2002cc}
  G.~Aarts and J.~M.~Martinez Resco,
  JHEP {\bf 0204} (2002) 053
  [hep-ph/0203177].
  
  
\bibitem{Arnold:2003zc}
  P.~B.~Arnold, G.~D.~Moore, L.~G.~Yaffe,
  JHEP {\bf 0305 } (2003)  051,
  [hep-ph/0302165].

  
\bibitem{Kovtun:2004de}
  P.~Kovtun, D.~T.~Son, A.~O.~Starinets,
  Phys.\ Rev.\ Lett.\  {\bf 94 } (2005)  111601,
  [hep-th/0405231].

\bibitem{Kovtun:2011np}
  P.~Kovtun, G.~D.~Moore and P.~Romatschke,
  Phys.\ Rev.\ D {\bf 84} (2011) 025006
  [arXiv:1104.1586 [hep-ph]].

\bibitem{Rebhan:2011vd}
  A.~Rebhan and D.~Steineder,
  Phys.\ Rev.\ Lett.\  {\bf 108} (2012) 021601
  [arXiv:1110.6825 [hep-th]].
 
\bibitem{Meyer:2007ic}
  H.~B.~Meyer,
  Phys.\ Rev.\ D {\bf 76} (2007) 101701
  [arXiv:0704.1801 [hep-lat]].
  
\bibitem{Meyer:2007dy}
  H.~B.~Meyer,
  Phys.\ Rev.\ Lett.\  {\bf 100} (2008) 162001
  [arXiv:0710.3717 [hep-lat]].
  

\bibitem{Meyer:2011gj}
  H.~B.~Meyer,
  Eur.\ Phys.\ J.\ A {\bf 47} (2011) 86
  [arXiv:1104.3708 [hep-lat]].

\bibitem{Brandt:2012jc}
  B.~B.~Brandt, A.~Francis, H.~B.~Meyer and H.~Wittig,
  arXiv:1212.4200 [hep-lat].


\bibitem{Huebner:2008as}
  K.~Huebner, F.~Karsch and C.~Pica,
  Phys.\ Rev.\ D {\bf 78} (2008) 094501
  [arXiv:0808.1127 [hep-lat]].
 
\bibitem{Iqbal:2009xz}
N.~Iqbal, H.~B.~Meyer,
JHEP {\bf 0911 } (2009)  029,
[arXiv:0909.0582].


\bibitem{Springer:2010mf}
  T.~Springer, C.~Gale, S.~Jeon, S.~H.~Lee,
  Phys.\ Rev.\  {\bf D82 } (2010)  106005.
  [arXiv:1006.4667].


\bibitem{Springer:2010mw}
  T.~Springer, C.~Gale, S.~Jeon,
  Phys.\ Rev.\  {\bf D82 } (2010)  126011,
  [arXiv:1010.2760].




\bibitem{Kajantie:2010nx}
  K.~Kajantie, M.~Vepsalainen,
  Phys.\ Rev.\  {\bf D83 } (2011)  066003,
  [arXiv:1011.5570].

\bibitem{Kajantie:2011nx}
  K.~Kajantie, M.~Krssak, M.~Vepsalainen and A.~Vuorinen,
  Phys.\ Rev.\ D {\bf 84} (2011) 086004
  [arXiv:1104.5352 [hep-ph]].

  
\bibitem{Gursoy:2007cb}
  U.~Gursoy and E.~Kiritsis,
  JHEP {\bf 0802} (2008) 032
  [arXiv:0707.1324 [hep-th]].

\bibitem{Jarvinen:2011qe}
  M.~Jarvinen and E.~Kiritsis,
  JHEP {\bf 1203} (2012) 002
  [arXiv:1112.1261 [hep-ph]].
  
\bibitem{Alho:2012mh}
  T.~Alho, M.~Jarvinen, K.~Kajantie, E.~Kiritsis and K.~Tuominen,
  arXiv:1210.4516 [hep-ph].
  
  
\bibitem{Laine:2011xm}
  M.~Laine, A.~Vuorinen and Y.~Zhu,
  JHEP {\bf 1109} (2011) 084
  [arXiv:1108.1259 [hep-ph]].

\bibitem{Zhu:2012be}
  Y.~Zhu and A.~Vuorinen,
  arXiv:1212.3818 [hep-ph].

  
\bibitem{Krssak} 
K.~Kajantie, M.~Krssak and A.~Vuorinen, In preparation.


\bibitem{Burnier:2011jq}
  Y.~Burnier, M.~Laine and L.~Mether,
  Eur.\ Phys.\ J.\ C {\bf 71} (2011) 1619
  [arXiv:1101.5534 [hep-lat]].
  
\bibitem{Burnier:2012ts}
  Y.~Burnier and M.~Laine,
  Eur.\ Phys.\ J.\ C {\bf 72} (2012) 1902
  [arXiv:1201.1994 [hep-lat]].
  
 

\bibitem{Burnier:2012ze}
 Y.~Burnier and M.~Laine,
 JHEP {\bf 1211} (2012) 086
 [arXiv:1210.1064 [hep-ph]].


  

  


\bibitem{Laine:2010tc}
  M.~Laine, M.~Vepsalainen, A.~Vuorinen,
  JHEP {\bf 1010 } (2010)  010,
  [arXiv:1008.3263].
  
\bibitem{Schroder:2011ht}
  Y.~Schroder, M.~Vepsalainen, A.~Vuorinen and Y.~Zhu,
  JHEP {\bf 1112} (2011) 035
  [arXiv:1109.6548 [hep-ph]].
  


\bibitem{Zoller:2012qv}
  M.~F.~Zoller and K.~G.~Chetyrkin,
  arXiv:1209.1516 [hep-ph].
  
  
\bibitem{CaronHuot:2009ns}
  S.~Caron-Huot,
  Phys.\ Rev.\  {\bf D79 } (2009)  125009,
  [arXiv:0903.3958]. 
    
\bibitem{Romatschke:2009ng}
  P.~Romatschke, D.~T.~Son,
  Phys.\ Rev.\  {\bf D80 } (2009)  065021,
  [arXiv:0903.3946].

\bibitem{Meyer:2010gu}
  H.~B.~Meyer,
  Phys.\ Rev.\  {\bf D82 } (2010)  054504,
  [arXiv:1005.2686].


\bibitem{Meyer:2010ii}
  H.~B.~Meyer,
  JHEP {\bf 1004} (2010) 099
  [arXiv:1002.3343 [hep-lat]].
  
 


  


\end{thebibliography}
\end{document}